\documentclass[conference]{IEEEtran}
\usepackage{cite}
\usepackage{amsmath,amssymb,amsfonts}
\usepackage{algorithmic}
\usepackage{graphicx}
\usepackage{textcomp}
\usepackage{xcolor}
\usepackage{hyperref}
\def\BibTeX{{\rm B\kern-.05em{\sc i\kern-.025em b}\kern-.08em
    T\kern-.1667em\lower.7ex\hbox{E}\kern-.125emX}}
\begin{document}

\title{Hierarchical Neural Surfaces for 3D Mesh Compression}

\author{Sai Karthikey Pentapati$^\star$, Gregoire Phillips$^\dagger$, Alan C. Bovik$^\star$

\\
\IEEEauthorblockA{$^\star$Laboratory of Image and Video Engineering, The University of Texas at Austin}
\IEEEauthorblockA{$^\dagger$Ericsson Research, Santa Clara, CA}
}

\maketitle

\begin{abstract}
Implicit Neural Representations (INRs) have been demonstrated to achieve state-of-the-art compression of a broad range of modalities such as images, videos, 3D surfaces, and audio.
Most studies have focused on building neural counterparts of traditional implicit representations of 3D geometries, such as signed distance functions.
However, the triangle mesh-based representation of geometry remains the most widely used representation in the industry, while building INRs capable of generating them has been sparsely studied.
In this paper, we present a method for building compact INRs of zero-genus 3D manifolds.
Our method relies on creating a spherical parameterization of a given 3D mesh - mapping the surface of a mesh to that of a unit sphere - then constructing an INR that encodes the displacement vector field defined continuously on its surface that regenerates the original shape.
The compactness of our representation can be attributed to its hierarchical structure, wherein it first recovers the coarse structure of the encoded surface before adding high-frequency details to it.
Once the INR is computed, 3D meshes of arbitrary resolution/connectivity can be decoded from it.
The decoding can be performed in real time while achieving a state-of-the-art trade-off between reconstruction quality and the size of the compressed representations.
\end{abstract}

\begin{IEEEkeywords}
3D Mesh Compression, Implicit Neural Representations, Neural Geometry Processing, Spherical Harmonics
\end{IEEEkeywords}

\section{Introduction}
Despite the emergence of many novel representations of 3D scenes in recent years, 3D meshes remain the most ubiquitously used because of their compact and continuous representation of geometry, ease of storing additional surface attributes (texture maps, normal maps, bump maps, etc), and ease of rendering with current graphics processing hardware.
With the advent of virtual, augmented, and mixed reality use-cases, compact storage and efficient distribution of 3D media have gained unprecedented importance.
Traditionally, research was focused on compressing meshes by either losslessly compressing the connectivity \cite{edgebreaker, tfan} and quantizing the geometry (as seen in industry standards like Google Draco\cite{draco}), or simplifying them by reducing their vertex count\cite{qslim, msqt}.
More recently, neural methods of compression have risen in prominence, with Implicit Neural Representations (INRs) \cite{siren} gaining significant traction.

INRs have multiple properties that make their study interesting for the use case of compression.
Firstly, the sizes of INRs, unlike pixel/voxel-based representations, are independent of resolution and only depend on the complexity of the underlying signal, and consequently, INRs can be queried at an infinite resolution.
While the bandwidth of the encoded function is limited by the capacity of the neural network used, the interpolation intrinsically performed by INRs when queried at arbitrary points tends to be more non-linear and can rely on a larger receptive field than interpolation of traditional discrete representations using methods such as bicubic, Lanczos \cite{lanczos}, barycentric, etc. where the receptive field is fixed by the kernel size, leading to more meaningful samplings. 
INRs are also infinitely differentiable, unlike continuous signals obtained by the aforementioned piecewise interpolation schemes applied on discrete data - a property that could be useful when direct application of continuous signal processing algorithms is needed, especially for discrete 3D geometry processing due to the notorious ``no free lunch'' theorem\cite{nflt}.
Additionally, due to per-instance training, INRs preserve high-frequency details much better than traditional compression techniques.
Owing to these advantages, many works have built INRs for that can infill missing sections of, interpolate, and compress different types of signals such as images \cite{imginr1, imginr2}, videos \cite{vidinr1, vidinr2}, signed distance functions (SDFs) of 3D shapes \cite{sdfinr1, sdfinr2, sdfinr3}, and radiance fields \cite{nerf1}.

\begin{figure*}
    \centering
    \includegraphics[width=0.95\linewidth]{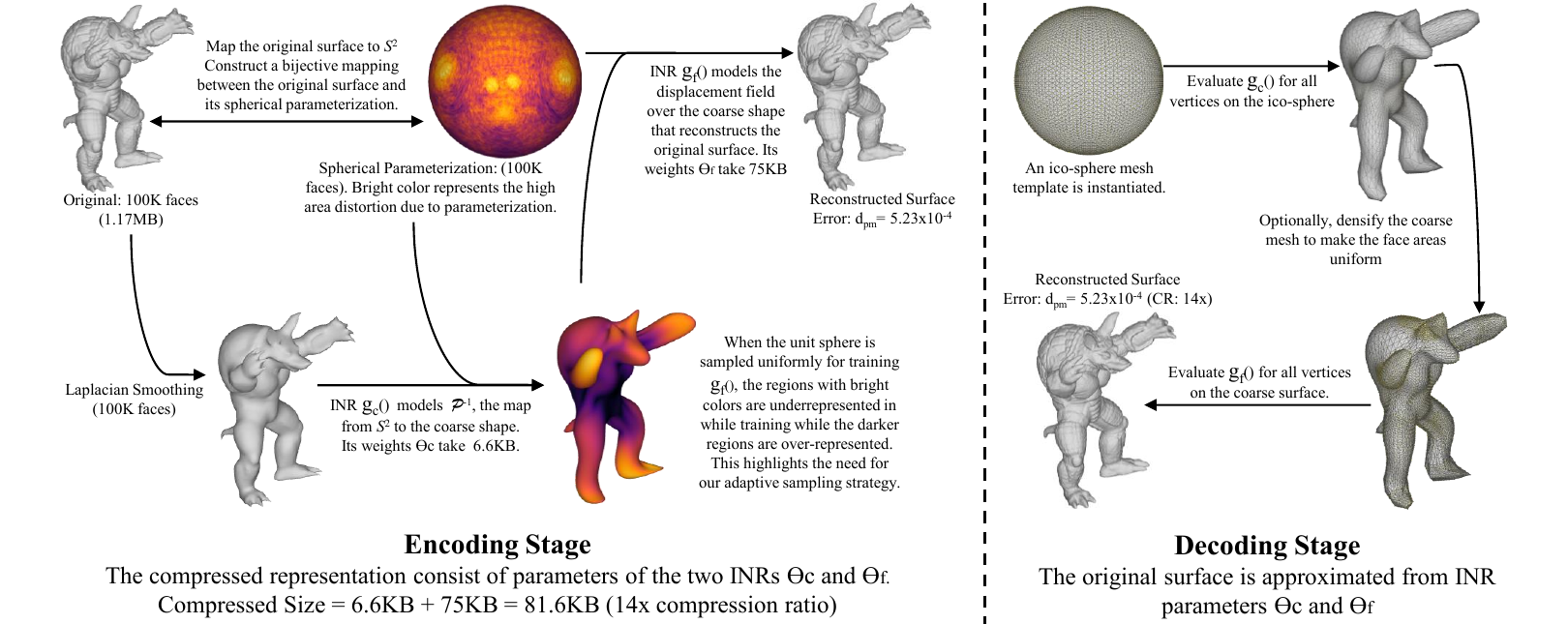}
    \caption{\textbf{Overview of our method for mesh compression:}
    A given surface is is represented as the inverse function of its spherical parameterization which is then modeled by a hierarchical INR.
    The methods yields high quality reconstructions even at high compression ratios.}
    \label{fig:overview}
\end{figure*}

Many recent works have focused on building compact neural representations of 3D shapes. ACORN\cite{acorn} and NGLOD\cite{nglod} employ octree-based partitioning of space for memory-efficient representation of the occupancy fields and signed distance functions, respectively. 
Implicit Displacement Fields (IDF)\cite{sdfinr2} encode the SDF as a composition of two INRs - the first encodes the SDF of the coarse shape while the second models a displacement field that adds details to the coarse surface. 
Although these methods yield impressive reconstruction quality, they are not memory optimal, as the neural networks must spend valuable learning capacity on points not on the surface. 
This is because the SDF/occupancy fields need to be well defined on any point in the ambient space around the 3D surface (octree-based/hierarchical representations mitigate this to some extent, but not fully). 
Additionally, these networks also have to be regularized to obey the Eikonal equations of SDFs by ensuring the magnitudes of their gradients are unity.
Reconstructing encoded surfaces from these INRs also tends to be slow, as they involve using an algorithm like Marching Cubes\cite{marchingcubes} to extract iso-surfaces from encoded SDFs.
Meshes extracted using Marching Cubes tend to suffer from topological inconsistencies, such as being non-watertight, and also contain discretization artifacts that are visually jarring.

To circumvent these issues, some methods have proposed storing a very coarse model of the surface explicitly as a part of the compressed representation and adding fine details to it using displacement fields encoded as INRs, avoiding the need to perform expensive iso-surface extraction from SDF-based representations.
Other methods tackle this issue by first parameterizing a surface to a predefined chart, such as a unit circle\cite{ncs} or a unit sphere\cite{sns}, then training an INR of the inverse function of the parameterization. 
Unlike SDF-based methods, these methods are capable of directly generating meshes in real time without topological artifacts.
The hybrid explicit-implicit representation of the former type of methods cannot be generalized to store surface attributes of meshes such as texture maps, normal maps etc, while the latter type of methods suffer warping artifacts, as it is often impossible to preserve the structure of triangles (area ratio, internal angles, etc.) when parameterized to a simpler shape.

Here we propose a method of compressing 3D meshes that belongs to the latter paradigm, but enriched with a hierarchical pipeline.
The hierarchical strategy allows us to implement a novel adaptive sampling technique during INR training, allowing us to accelerate it and improve reconstruction quality by mitigating warping artifacts to a significant extent, allowing our method to obtain state-of-the-art compression of 3D meshes. 
Our representation is also infinitely differentiable and can be extended to compress surface maps, unlike current leading mesh compression methods, which employ hybrid explicit-implicit structures.

\section{Method}
\subsection{Overview} 
\label{subsec:overview}

Figure \ref{fig:overview} illustrates the overall pipeline of our method. 
Formally, consider a 3D mesh surface $\mathcal{M}(V,F)\subset\mathbb{R}^3$ with vertices $V\in\mathbb{R}^3$ and faces $F$.
Its encoding process can be broken down into the following steps:
\begin{enumerate}
    \item Generate the spherical parametrization $\mathcal{P}$ of $\mathcal{M}$ by mapping all vertices $\mathcal{P}(V) = V_s$ injectively to the surface a unit sphere $(V_s\in)S^2$ while maintaining the same connectivity $F$ and its topology to obtain $\mathcal{M}_s(V_s, F)$. We use the algorithm proposed by Praun and Hoppe\cite{spr} to perform this mapping.
    \item Extend this injection to a bijection $\mathcal{P}:\mathcal{M}\leftrightarrow\mathcal{M}_s$ by first representing arbitrary surface points on $\mathcal{M}$ as a barycentric interpolation of the vertices of the triangle that contains it, then using the same barycentric coordinates to interpolate the vertices of the same face in $\mathcal{M}_s$.
    \item Apply $c$ iterations of Laplacian smoothing\cite{lapl} to $\mathcal{M}$ to generate its coarse version $\mathcal{M}_c$. 
    It is important to note that $\mathcal{M}, \mathcal{M}_s\text{, and }\mathcal{M}_c$ share the same connectivity and topology ($F$).
    As such, a bijection $\mathcal{P}_c:\mathcal{M}_c\leftrightarrow\mathcal{M}_s$ can also be defined without explicitly recomputing the spherical parameterization of $\mathcal{M}_c$.
    \item The inverse of the parameterization $\mathcal{P}^{-1}_c: \mathcal{M}_s\to\mathcal{M}_c$ is approximated by training a tiny multilayer perceptron \cite{mlp} $q_c(\cdot)$ with parameters $\theta_c$ in a manner similar to \cite{sns}. 
    Thus an INR of $\mathcal{P}^{-1}_c$ is built, which can generate $\mathcal{M}^*_c$, an approximation of $\mathcal{M}_c$, from $S^2$.
    See Section \ref{subsec:coarse} for more details.
    \item Finer high-resolution details are added to $\mathcal{M}^*_c$ using a second MLP $q_f(\cdot)$ with parameters $\theta_f$ that forms an INR that encodes the displacement field $\mathcal{D}:\mathcal{M}^*_c\to\mathbb{R}^3$ such that $\mathcal{M}^*:=\mathcal{M}_c^*+\mathcal{D}(\mathcal{M}_c^*)$ closely approximates $\mathcal{M}$.
    Its training using our adaptive sampling is detailed in Section \ref{subsec:fine}.
\end{enumerate}
The parameters of the two INRs, $\theta_c$ and $\theta_f$, form our compressed representation. 
Optionally, these parameters can be pruned, quantized, and entropy coded to obtain further compression. 
The connectivity information $F$ is not included in the compressed representation.
The decoder reconstructs a mesh from this hierarchical INR in the following manner:
\begin{enumerate}
    \item Initialize an icosphere mesh $\mathcal{I}_k(V_k, F_k)$ of level $k$.  
    Starting from a unit icosahedron, an icosphere mesh of level $k$ can be created by iteratively subdividing and projecting the newly introduced vertices to the unit sphere $k$ times.
    \item Project all vertices $V_k$ to the coarse surface by evaluating $V^*_c = q_c(V_k)$. 
    This yields $\mathcal{M}^*_c(V^*_c, F_k)$.
    \item Optionally, identify if some faces of $\mathcal{M}^*_c$ are significantly larger than the rest.
    We heuristically identify all the faces whose areas are more than four times the median face area.
    The corresponding faces in $\mathcal{I}_k$ are subdivided, and the newly introduced vertices are pushed through $q_c$ to obtain a version of $\mathcal{M}^*_c$ with more uniform face areas.
    This process can be repeated to obtain the reconstructed topology $F^*_k$ until no faces are identified by the heuristic.
    \item Evaluate the displacement field at all vertices of the reconstructed coarse mesh, and apply it to the same vertices: $V^* = q_f(V^*_s) + V^*_s$. 
    The obtained reconstruction $\mathcal{M}^*(V^*, F^*_k)$ approximates the original mesh $\mathcal{M}(V,F)$.
\end{enumerate}

The original mesh $\mathcal{M}$  may be accompanied by an attribute map $\mathcal{S}(p):\mathcal{M}\to\mathbb{R}^d, p\in\mathcal{M}$ such as a texture, normal, or bump map, whose encoding can be performed similarly but is omitted from the above formulation for brevity.

\subsection{Coarse INR Training}
\label{subsec:coarse}
The process of training an INR that reconstructs the coarse surface largely follows \cite{sns}. 
First, points are sampled uniformly on $\mathcal{M}_s$.
This can be trivially performed by first assigning a probability to each face in $\mathcal{M}_s$ which is proportional to its area, then drawing faces according to the resulting distribution, and finally randomly selecting points on those faces.
These sampled points are normalized to unit length to ensure they lie on $S^2$.
(This step is needed because the sampled points, although are already very close to $S^2$ as $\mathcal{M}_s$ is the spherical parameterization of $\mathcal{M}$, may not lie exactly on it due to the discrete mesh nature of $\mathcal{M}_s$).
Using the same barycentric interpolation trick as the second step of the encoding process described in Section \ref{subsec:overview}, $\mathcal{P}^{-1}_c$ is evaluated for all the sampled points.
These evaluated samples serve as ground truths for training the INR $q_c$ by minimizing the expectation $\text{E}_{p\in S^2}(||q_c(p)-\mathcal{P}_c^{-1}(p)||_2^2)$, where $p$ are sampled uniformly from $S^2$.

\subsection{Fine INR Training}
\label{subsec:fine}
Adopting a batch sampling strategy similar to coarse training poses a limitation.
Sampling $\mathcal{M}_s$ uniformly and evaluating $q_c$ for them does not yield samples that are uniformly distributed on $\mathcal{M}^*_c$.
As a result, uniformly sampling $\mathcal{M}_s$ for training $q_f$ leads to inefficient distribution of the INR's model capacity.
To mitigate this issue, we introduce an adaptive sampling strategy that is enabled by the hierarchical structure of our representation.

\noindent\textbf{Prerequisite - The Metric Tensor:} As $q_c$ is a map from $S^2\to\mathbb{R}^3$, and $S^2$ can be further parameterized using only two variables - the azimuthal angle $u$ and the polar angle $v$ - via a composition of functions, $q(u,v):= q_c(S^2(u,v))$, a map from $\mathbb{R}^2$ to $\mathbb{R}^3 $ can be obtained.
The metric tensor $\text{I}(u,v)$ (also known as the first fundamental form) of this mapping is defined at each point in the parameter space as shown below:
\begin{equation}
    \text{I}(u,v) = \begin{bmatrix}
        \frac{\partial q}{\partial u}\cdot\frac{\partial q}{\partial u} & \frac{\partial q}{\partial u}\cdot\frac{\partial q}{\partial v} \\ \frac{\partial q}{\partial u}\cdot\frac{\partial q}{\partial v} & \frac{\partial q}{\partial v}\cdot\frac{\partial q}{\partial v}
    \end{bmatrix}
\end{equation}
Then the distortion ratio $d(u,v) = \sqrt{\text{det}(\text{I}(u,v))/\text{sin}^2(v)}$ gives the factor by which an infinitesimal area on $S^2$ at $(u,v)$ is multiplied when pushed through the map $q$.
By virtue of representing this mapping using the neural network $q_c$, it is continuously differentiable, and its metric tensor can be evaluated at any arbitrary point. 

Finally, the batches for training $q_f$ can be generated in the following manner:
\begin{enumerate}
    \item An icosphere of a relatively low level is instantiated. 
    \item The distortion ratio $d$ is evaluated at each of its face centers.
    These two steps are performed only once before the start of the training.
    \item Samples $p$ are drawn from the icosphere with face probabilities being proportional to their distortion ratios.
    \item $\mathcal{P}^{-1}(p)$ is obtained for these points using the same barycentric interpolation trick to obtain the corresponding points lying on the original mesh $\mathcal{M}$.
    \item The displacement field $\mathcal{D}(p) = \mathcal{P}^{-1}(p)-q_c(p)$ at these points is evaluated which is used as a training target.
\end{enumerate}
Finally, $q_f$ is trained by minimizing $\text{E}_{p\in S^2}(||q_f(q_c(p))-\mathcal{D}(p)||_2^2)$, where $p$ are sampled from $S^2$ using the aforementioned strategy. 
The parameters of $q_c$ are not optimized in this stage. 
This data sampling strategy closely mimics the mesh reconstruction process performed by the decoder and consequently enhances the compression outcomes.

\begin{table}
\caption{Mean $d_{pm}\times10^4$ and $d_n (^\circ)$ are shown for 15 meshes from Thingi10K dataset compressed to different target sizes}
\resizebox{\columnwidth}{35px}{
\begin{tabular}{|c|c|c|c|c|}
\hline
\textbf{Method}     & \textbf{50KB} & \textbf{85KB} & \textbf{165KB} & \textbf{260KB} \\ \hline
\textbf{Ours}       & 6.56 / 7.12   & 4.80 / 5.69   & 3.11 / 3.20    & 2.41 / 3.23    \\
\textbf{QNDF}       & 9.12 / 10.67  & 6.92 / 6.52   & 3.93 / 3.90    & 2.43 / 3.34    \\
\textbf{SNS (Qnt.)} & 16.31 / 13.89 & 11.77 / 9.15  & 7.07 / 7.41    & 5.14 / 5.18    \\ \hline
\textbf{Ours (NQ)}  & 13.11 / 9.91  & 9.05 / 7.08   & 5.98 / 4.87    & 5.42 / 4.69    \\
\textbf{QNDF(NQ)}   & 15.75 / 13.26 & 11.98 / 7.99  & 6.88 / 5.41    & 5.30 / 4.69    \\
\textbf{SNS}        & 20.12 / 19.31 & 14.42 / 12.54 & 10.40 / 9.43   & 8.01 / 7.87    \\ \hline
\textbf{QS-DRC}     & 24.41 / 17.92 & 15.05 / 11.99 & 7.23 / 4.70    & 3.18 / 3.44    \\ \hline
\end{tabular}
}
\label{tab:results}
\end{table}

\subsection{Model Architecture}

We model both $q_c$ and $q_f$ using MLPs with residual connections and SiLU activations\cite{gelu}. 
The input to $q_c$ is a 3D coordinate $(x,y,z) \in S^2$ and the output is the estimated 3D position on the coarse mesh $\mathcal{M}^*_c$.
On the other hand, instead of passing 3D coordinates on $\mathcal{M}^*_c$ to $q_f$ directly, we first positionally encode them in the same manner as \cite{nerf1} to overcome the spectral bias of neural networks\cite{bias}.
It is unnecessary, and even undesirable positionally encode the inputs to $q_c$ as (a) it does not have to encode a very high frequency signal; (b) makes it prone to overfitting; and (c) necessitates the increase in the parameters in the first layer of the MLP, consequently the size of the compressed representation.
The number of layers and the hidden units per layer are the hyperparameters that determine the compression ratio. 
We detail the variation in reconstruction quality with model size in Section \ref{expt}.
For improved compressed outcomes, we quantize the parameters of the MLP to 16-bit floats post-training, similar to \cite{nmc}.

\section{Experiments}
\label{expt}

\begin{figure}[ht]
    \centering
    \includegraphics[width=\linewidth]{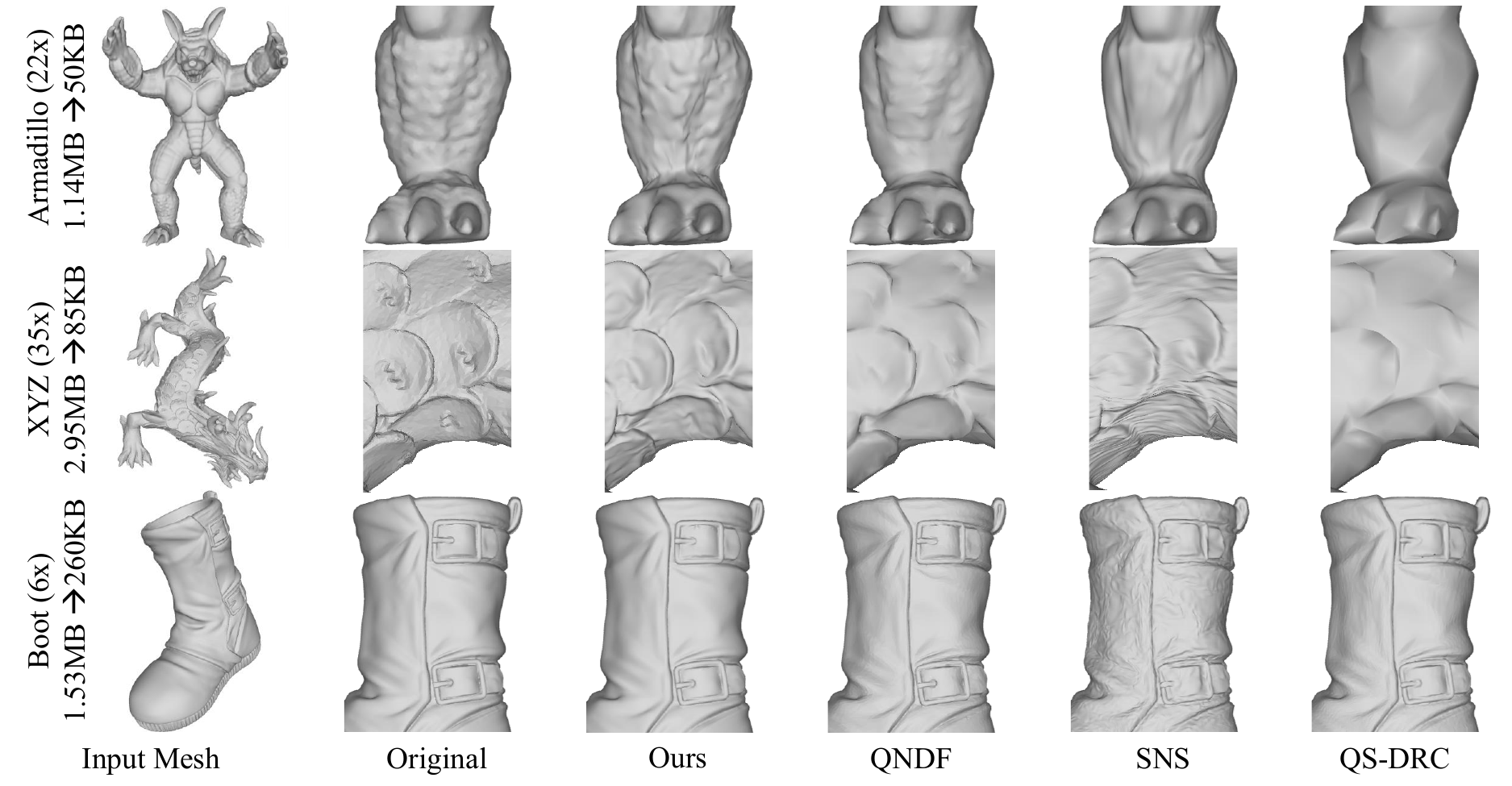}
    \vspace{-15px}
    \caption{Our method accurately reconstructs the originals over a broad range of bitrates. \texttt{QNDF} has visible quantization artifacts, \texttt{SNS} suffers warping artifacts, while \texttt{QS-DRC} produces very low resolution meshes}
    \label{fig:qual}
\end{figure}

We evaluated our method on 15 watertight zero-genus meshes from Thingi10k dataset\cite{thingi} representing a wide variety of complex geometries and varying vertex counts.
We compressed each of them to four sizes (50KB, 85KB, 165KB, and 260KB) and compared them to the following baselines:
\begin{enumerate}
    \item Quantized Neural Displacement Fields (\texttt{QNDF}) \cite{nmc}: A hybrid explicit-implicit INR that models the displacement field on the decimated version of the surface to be encoded - the current state-of-the-art INR for 3D mesh compression.
    \item Spherical Neural Surfaces (\texttt{SNS}) \cite{sns}: The non-hierarchical analogue of our representation.
    We replaced the activation function and scheduler used in their method with our choices, which increases its performance by a small margin.
    \item Q-SLIM \cite{qslim} + Google Draco with quantization \cite{draco}: A competitive non-INR-based baseline for lossy compression that can encode in interactive time frames.
\end{enumerate}
Results of all INR-based methods are presented with and without quantization of model parameters.
The reconstruction quality of all methods and bitrates was measured in terms of the mean point-to-mesh error and the mean normal error, and is displayed in Table \ref{tab:results}.
Qualitative comparisons are also shown in Figure \ref{fig:qual}.
Our method yields greater improvement over existing methods for high compression ratios. 
For lower compression ratio, we find non-INR based methods to yield comparable reconstruction quality while running in real time.

\subsection{Training Details}
For all bitrates, our MLP $q_c$ contains 20 hidden layers with 12 hidden nodes in each. 
When its parameters are quantized to 16 bits, it takes up $6.6$KB, else $12.2$KB.
Similarly, we performed positional encoding of 3D coordinates up to 10 frequency levels before feeding them to the second MLP $q_f$.
The batch size for training both $q_c$ and $q_f$ was set to 2048. 
$q_c$ was trained for 50,000 iterations ($\sim8$ minutes wall clock time), while $q_f$ was trained for 15,000 iterations using the AdamW optimizer with cosine annealing with a learning rate of $10^{-3}$.
Our choices of hyperparameters of $q_f$ for different bitrates models, training time and GPU usage, decoding time and GPU usage when using a level-8 icosphere template, and as input are listed in Table \ref{tab:params}.
The total encoding time per shape ranges from 25 to 35 minutes using our method, which is comparable to other INR-based compression methods, such as \texttt{QNDF} and \cite{ngf}. 
Our hierarchical approach also makes our method faster than \texttt{SNS}, which takes 8-10 hours to train on average.
While encoding meshes using our method is time-consuming, a mesh with 327K vertices can be decoded in under 100ms.
This is magnitudes faster than SDF-based neural representations (2-4 minutes per shape), and similar to INR-based mesh compression methods such as $\texttt{QNDF}$.
All our experiments were performed on an NVIDIA RTX 2070 GPU.

The encoded mesh results and the code for our method can be found in \href{https://github.com/karthikey97/HierarchicalNeuralSurfaces/tree/decoder}{our project page}.

\begin{figure}[h]
    \centering
    \includegraphics[width=\linewidth]{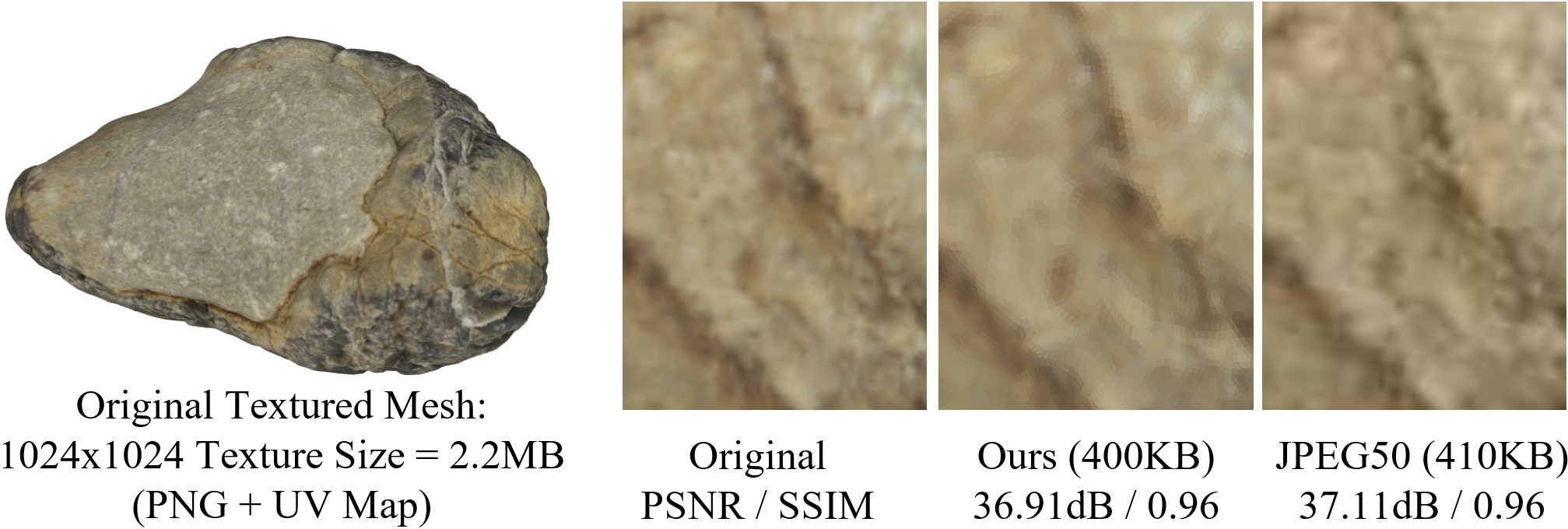}
    \vspace{-15px}
    \caption{Our method can be successfully extended for compressing surface maps like textures.}
    \label{fig:txtrcomp}
\end{figure}

\begin{table}[h]
\caption{Training parameters for various target sizes}
\resizebox{\columnwidth}{19px}{
\begin{tabular}{|c|c|c|c|c|c|}
\hline
Compressed Size & \# Hidden Layers & \# Hidden Units & Train Time & VRAM Use & Decode Time \\ \hline
50 KB           & 18               & 36              & 14 mins    & 320MB     & 230ms       \\
85KB            & 20               & 44              & 14 mins    & 360MB     & 280ms       \\
165KB           & 24               & 58              & 17 mins    & 480MB     & 330ms       \\
260KB           & 28               & 68              & 20 mins    & 600MB     & 440ms       \\ \hline
\end{tabular}
}
\label{tab:params}
\end{table}

\subsection{Additional Experiments}
An example of encoding a texture map of size $1024\times1024$ is shown in Figure \ref{fig:txtrcomp}.
The mesh with the reconstructed texture is rendered to an image, and its PSNR and SSIM\cite{ssim} with respect to the image rendered using the original texture are also denoted.
Our method performs comparably to JPEG compression with $qp=50$ slightly lower memory.
Note that the memory required to store the JPEG texture includes the memory required to store its corresponding UV parameterization as well, whereas our representation has no such need.
Additionally, as our representation is essentially the same as \cite{sns} but with a hierarchical structure, we can perform differential operations on it in similar manner.

\section{Conclusion}
In this paper, we presented a method of building compact INRs to represent 3D meshes that yields state-of-the-art compression outcomes.
Unlike previous 3D mesh compression methods, we also demonstrated the encoding of surface attributes.
Additionally, due to the fully differentiable nature of our representation, we could also compute differential quantities of the encoded surface directly from the representation.

Due to the need for per-instance training, our encoding times are slow, hindering this method for real-time use cases.
Future research in this direction is of extreme interest, and extending hybrid explicit-implicit pretrained encoders such as \cite{cofie} to spherical formulations could be a promising direction.
Additionally, we only demonstrated our method to zero-genus surfaces.
It should be possible to extend it to arbitrary genera by choosing appropriate template shapes and parametrization algorithms, but we leave it for future work.

We believe our work demonstrates the efficacy and flexibility of neural representations for compressing 3D geometries and motivates further research to accelerate and improve them for feasible adaptation in industrial workflows.


\begin{thebibliography}{00}
\bibitem{siren} Vincent Sitzmann, Julien N. P. Martel, Alexander W. Bergman, David B. Lindell, and Gordon Wetzstein. 2020. Implicit neural representations with periodic activation functions. In Proceedings of the 34th International Conference on Neural Information Processing Systems.
\bibitem{lanczos} Duchon, C. E., 1979: Lanczos Filtering in One and Two Dimensions. J. Appl. Meteor. Climatol., 18, 1016–1022
\bibitem{imginr1} Yannick Strümpler, Janis Postels, Ren Yang, Luc Van Gool, and Federico Tombari. 2022. Implicit Neural Representations for Image Compression. In Computer Vision – ECCV 2022: 17th European Conference.
\bibitem{imginr2} V. Saragadam et al., ``WIRE: Wavelet Implicit Neural Representations," 2023 IEEE/CVF Conference on Computer Vision and Pattern Recognition (CVPR), Vancouver, BC, Canada, 2023, pp. 18507-18516, doi: 10.1109/CVPR52729.2023.01775.
\bibitem{vidinr1} Z. Chen et al., ``VideoINR: Learning Video Implicit Neural Representation for Continuous Space-Time Super-Resolution," 2022 IEEE/CVF Conference on Computer Vision and Pattern Recognition (CVPR), pp. 2037-2047, doi: 10.1109/CVPR52688.2022.00209.
\bibitem{vidinr2} H. Chen, M. Gwilliam, S. -N. Lim and A. Shrivastava, ``HNeRV: A Hybrid Neural Representation for Videos," 2023 IEEE/CVF Conference on Computer Vision and Pattern Recognition (CVPR), Vancouver, BC, Canada, 2023, pp. 10270-10279, doi: 10.1109/CVPR52729.2023.00990.
\bibitem{sdfinr1} J. J. Park, P. Florence, J. Straub, R. Newcombe and S. Lovegrove, ``DeepSDF: Learning Continuous Signed Distance Functions for Shape Representation," 2019 IEEE/CVF Conference on Computer Vision and Pattern Recognition, 2019, pp. 165-174, doi: 10.1109/CVPR.2019.00025.
\bibitem{sdfinr2} Yifan, Wang, Rahmann, Lukas \& Sorkine-Hornung, Olga. (2021). Geometry-Consistent Neural Shape Representation with Implicit Displacement Fields. 10.48550/arXiv.2106.05187. 
\bibitem{sdfinr3} V. Saragadam et al. 2022. MINER: Multiscale Implicit Neural Representation. In Computer Vision – ECCV 2022: 17th European Conference, Tel Aviv, Israel, October 23–27, 2022, Proceedings, Part XXIII. 318–333. https://doi.org/10.1007/978-3-031-20050-2\_19
\bibitem{nerf1} Ben Mildenhall, Pratul P. Srinivasan, Matthew Tancik, Jonathan T. Barron, Ravi Ramamoorthi, and Ren Ng. 2021. NeRF: representing scenes as neural radiance fields for view synthesis. Commun. ACM 65, 1 (January 2022), 99–106. https://doi.org/10.1145/3503250
\bibitem{acorn} Julien N. P. Martel, David B. Lindell, Connor Z. Lin, Eric R. Chan, Marco Monteiro, and Gordon Wetzstein. 2021. Acorn: adaptive coordinate networks for neural scene representation. ACM Trans. Graph. 40, 4, Article 58. https://doi.org/10.1145/3450626.3459785
\bibitem{nglod} T. Takikawa et al., ``Neural Geometric Level of Detail: Real-time Rendering with Implicit 3D Shapes," 2021 IEEE/CVF Conference on Computer Vision and Pattern Recognition (CVPR), Nashville, TN, USA, 2021, pp. 11353-11362, doi: 10.1109/CVPR46437.2021.01120.
\bibitem{marchingcubes} William E. Lorensen and Harvey E. Cline. 1987. Marching cubes: A high resolution 3D surface construction algorithm. SIGGRAPH Comput. Graph. 21, 4 (July 1987), 163–169. https://doi.org/10.1145/37402.37422
\bibitem{ngf} Venkataram Edavamadathil Sivaram, Tzu-Mao Li, and Ravi Ramamoorthi. 2024. Neural Geometry Fields For Meshes. In ACM SIGGRAPH 2024 Conference Papers (SIGGRAPH '24). Article 29, 1–11. https://doi.org/10.1145/3641519.3657399
\bibitem{nmc} Sai Karthikey Pentapati, Gregoire Phillips, and Alan C. Bovik. 2025. Mesh Compression with Quantized Neural Displacement Fields. Computer Graphics Forum (2025). DOI:https://doi.org/10.1111/cgf.70074
\bibitem{ncs} L. Morreale, N. Aigerman, P. Guerrero, V. G. Kim and N. J. Mitra, ``Neural Convolutional Surfaces," 2022 IEEE/CVF Conference on Computer Vision and Pattern Recognition (CVPR), New Orleans, LA, USA, 2022, pp. 19311-19320, doi: 10.1109/CVPR52688.2022.01873.
\bibitem{sns} Romy Williamson and Niloy J. Mitra. 2025. Neural Geometry Processing via Spherical Neural Surfaces. Computer Graphics Forum (2025). DOI:https://doi.org/10.1111/cgf.70021
\bibitem{nflt} Max Wardetzky, Saurabh Mathur, Felix Kälberer, and Eitan Grinspun. 2007. Discrete laplace operators: no free lunch. In Proceedings of the fifth Eurographics symposium on Geometry processing (SGP '07). Eurographics Association, Goslar, DEU, 33–37.
\bibitem{edgebreaker} J. Rossignac, ``Edgebreaker: connectivity compression for triangle meshes," in IEEE Transactions on Visualization and Computer Graphics, vol. 5, no. 1, pp. 47-61, Jan.-March 1999, doi: 10.1109/2945.764870.
\bibitem{tfan} K. Mamou, T. Zaharia and F. Prêteux, ``A Triangle-Fan-based approach for low complexity 3D mesh compression," 2009 16th IEEE International Conference on Image Processing (ICIP), Cairo, Egypt, 2009, pp. 3513-3516, doi: 10.1109/ICIP.2009.5414075.
\bibitem{draco} Galligan F., Hemmer M., Strava O., Zhang F., Brettle J.: Google/draco: a library for compressing and decompressing 3d geometric meshes and point clouds, 2018. URL: https://google.github.io/draco/
\bibitem{qslim} Michael Garland and Paul S. Heckbert. 1997. Surface simplification using quadric error metrics. In Proceedings of the 24th annual conference on Computer graphics and interactive techniques (SIGGRAPH '97). 209–216. https://doi.org/10.1145/258734.258849.
\bibitem{msqt} G. Zhou, S. Yuan and S. Luo, ``Mesh Simplification Algorithm Based on the Quadratic Error Metric and Triangle Collapse," in IEEE Access, vol. 8, pp. 196341-196350, 2020, doi: 10.1109/ACCESS.2020.3034075.
\bibitem{spr} Emil Praun and Hugues Hoppe. 2003. Spherical parametrization and remeshing. ACM Trans. Graph. 22, 3 (July 2003), 340–349. https://doi.org/10.1145/882262.882274
\bibitem{lapl} Mathieu Desbrun, Mark Meyer, Peter Schröder, and Alan H. Barr. 1999. Implicit fairing of irregular meshes using diffusion and curvature flow. In Proceedings of the 26th annual conference on Computer graphics and interactive techniques (SIGGRAPH '99), https://doi.org/10.1145/311535.311576
\bibitem{mlp} Rumelhart, D., Hinton, G. \& Williams, R. Learning representations by back-propagating errors. Nature 323, 533–536 (1986). https://doi.org/10.1038/323533a0
\bibitem{gelu} Dan Hendrycks, Kevin Gimpel. 2016. Gaussian Error Linear Units, https://doi.org/10.48550/arXiv.1606.08415
\bibitem{bias} Rahaman, N., Baratin, A., Arpit, D., Draxler, F., Lin, M., Hamprecht, F., Bengio, Y. \& Courville, A.. (2019). On the Spectral Bias of Neural Networks. Proceedings of the 36th International Conference on Machine Learning, in Proceedings of Machine Learning Research 97:5301-5310.
\bibitem{cofie} Jiang, Hanwen and Yang, Haitao and Pavlakos, Georgios and Huang, Qixing (2025). CoFie: Learning Compact Neural Surface Representations with Coordinate Fields. arXiv preprint arXiv:2406.03417
\bibitem{adam} Loshchilov I, Hutter F. Decoupled weight decay regularization. arXiv preprint arXiv:1711.05101. 2017 Nov 14.
\bibitem{canneal} Loshchilov I, Hutter F. SGDR: Stochastic Gradient Descent with Warm Restarts. arXiv preprint arXiv:1608.03983. 2016 Aug 13.
\bibitem{ssim} Zhou Wang, A. C. Bovik, H. R. Sheikh and E. P. Simoncelli, ``Image quality assessment: from error visibility to structural similarity," in IEEE Transactions on Image Processing, vol. 13, no. 4, pp. 600-612, April 2004, doi: 10.1109/TIP.2003.819861.
\bibitem{thingi} Zhou, Qingnan and Jacobson, Alec, Thingi10K: A Dataset of 10,000 3D-Printing Models, 2016, arXiv preprint arXiv:1605.04797.


\end{thebibliography}
\end{document}